# Microreboot – A Technique for Cheap Recovery

George Candea, Shinichi Kawamoto, Yuichi Fujiki, Greg Friedman, Armando Fox
*Computer Systems Lab, Stanford University*
{candea,skawamo,fjk,gregjf,fox}@cs.stanford.edu

**Abstract**

A significant fraction of software failures in large-scale Internet systems are cured by rebooting, even when the exact failure causes are unknown. However, rebooting can be expensive, causing nontrivial service disruption or downtime even when clusters and failover are employed. In this work we separate process recovery from data recovery to enable microrebooting – a fine-grain technique for surgically recovering faulty application components, without disturbing the rest of the application.

We evaluate microrebooting in an Internet auction system running on an application server. Microreboots recover most of the same failures as full reboots, but do so an order of magnitude faster and result in an order of magnitude savings in lost work. This cheap form of recovery engenders a new approach to high availability: microreboots can be employed at the slightest hint of failure, prior to node failover in multi-node clusters, even when mistakes in failure detection are likely; failure and recovery can be masked from end users through transparent call-level retries; and systems can be rejuvenated by parts, without ever being shut down.

## 1 Introduction

In spite of ever-improving development processes and tools, all production-quality software still has bugs; most of the bugs that escape testing are difficult to track down and resolve, and they take the form of Heisenbugs, race conditions, resource leaks, and environment-dependent bugs [14, 36]. Moreover, up to 80% of bugs that manifest in production systems have no fix available at the time of failure [43]. Fortunately, it is mostly application-level failures that bring down enterprise-scale software [32, 13, 35, 36], while the underlying platform (hardware and operating system) is reliable, by comparison. This is in contrast to smaller-scale systems, such as desktop computers, where hardware and operating system-level problems are still significant causes of downtime.

When failure strikes large-scale software systems, such as the ones found in Internet services, operators cannot afford to run real-time diagnosis. Instead, they focus on bringing the system back up by all means, and then do the diagnosis later. Our challenge is to find a simple, yet practical and effective approach to managing failure in large, complex systems, an approach that is accepting of the fact that bugs in application software will not be eradicated any time soon. The results of several studies [39, 19, 34, 12] and experience in the field [5, 35, 24] suggest that many failures can be successfully recovered by rebooting, even when the failure's root cause is unknown. Not surprisingly, today's state of the art in achieving high availability for Internet clusters involves circumventing a failed node through failover, rebooting the failed node, and subsequently reintegrating the recovered node into the cluster.

Reboots provide a high-confidence way to reclaim stale or leaked resources, they do not rely on the correct functioning of the rebooted system, they are easy to implement and automate, and they return the software to its start state, which is often its best understood and best tested state. Unfortunately, in some systems, unexpected reboots can result in data loss and unpredictable recovery times. This occurs most frequently when the software lacks clean separation between data recovery and process recovery. For example, performance optimizations, such as write-back buffer caches, open a window of vulnerability during which allegedly-persistent data is stored only in volatile memory; an unexpected crash and reboot could restart the system's processes, but buffered data would be lost.

This paper presents a practical recovery technique we call *microreboot* – individual rebooting of fine-grain application components. It can achieve many of the same benefits as whole-process restarts, but an order of magnitude faster and with an order of magnitude less lost work. We describe here general conditions necessary for microreboots to be safe: well-isolated, stateless components, that keep all important application state in specialized state stores. This way, data recovery is completely separated from (reboot-based) application recovery. We also describe a prototype microrebootable system we used in evaluating microreboot-based recovery.

The low cost of microrebooting engenders a new approach to high availability, in which microrebooting is always attempted first, as front-line recovery, even when failure detection is prone to false positives or when the failure is not known to be microreboot-curable. If the microreboot does not recover the system, but some other subsequent recovery action does, the recovery time added by the initial microreboot attempt is negligible. In multi-node clusters, a microreboot may be preferable even over node failover, because it avoids overloading non-failed nodes and preserves in-memory state. Being minimally-disruptive allows microreboots to rejuvenate a system by parts without shutting down; it also allows transparent call-level retries to mask a microreboot from end users.

The rest of this paper describes, in Section 2, a design for microrebootable software and, in Section 3, a prototype implementation. Sections 4 and 5 evaluate the prototype's recovery properties using fault injection and a realistic workload. Section 6 describes a new, simpler approach to failure management that is brought about by cheap recovery. Section 7 discusses limitations of microrebooting, and Section 8 presents a roadmap for generalizing our approach beyond the implemented prototype. Section 9 presents related work, and Section 10 concludes.

## 2 Designing Microrebootable Software

Workloads faced by Internet services often consist of many relatively short tasks, rather than long-running ones. This affords the opportunity for recovery by reboot, because any work-in-progress lost due to rebooting represents a small fraction of requests served in a day. We set out to optimize large-scale Internet services for frequent, fine-grain rebooting, which led to three design goals: fast and correct component recovery, strongly-localized recovery with minimal impact on other parts of the system, and fast and correct reintegration of recovered components.

In earlier work we introduced and motivated crash-only software [9] – programs that can be safely crashed in whole or by parts and recover quickly every time. The high-level recipe for building such systems is to structure them as a collection of small, well-isolated components, to separate important state from the application logic and place it in dedicated state stores, and to provide a framework for transparently retrying requests issued to components that are temporarily unavailable (e.g., because they are microrebooting). Here we summarize the main points of our crash-only design approach.

**Fine-grain components**: Component-level reboot time is determined by how long it takes for the underlying platform to restart a target component and for this component to reinitialize. A microrebootable application therefore aims for components that are as small as possible, in terms of program logic and startup time. (There are many other benefits to this design, which is why it is favored for large-scale Internet software.) While partitioning a system into components is an inherently system-specific task, developers can benefit from existing component-oriented programming frameworks, as will be seen in our prototype.

**State segregation**: To ensure recovery correctness, we must prevent microreboots from inducing corruption or inconsistency in application state that persists across microrebooting. The inventors of transactional databases recognized that segregating recovery of persistent data from application logic can improve the recoverability of both the application and the data that must persist across failures. We take this idea further and require that microrebootable applications keep *all* important state in dedicated state stores located outside the application, safeguarded behind strongly-enforced high-level APIs. Examples of such state stores include transactional databases and session state managers [26].

Aside from enabling safe microreboots, the complete separation of data recovery from application recovery generally improves system robustness, because it shifts the burden of data management from the often-inexperienced application writers to the specialists who develop state stores. While the number of applications is vast and their code quality varies wildly, database systems and session state stores are few and their code is consistently more robust. In the face of demands for ever-increasing feature sets, application recovery code that is both bug-free and efficient will likely be increasingly elusive, so data/process separation could improve dependability by making process recovery simpler. The benefits of this separation can often outweigh the potential performance overhead.

**Decoupling**: Components must be loosely coupled, if the application is to gracefully tolerate a microreboot (μRB). Therefore, components in a crash-only system have well-defined, well-enforced boundaries; direct references, such as pointers, do not span these boundaries. If cross-component references are needed, they should be stored outside the components, either in the application platform or, in marshalled form, inside a state store.

**Retryable requests**: For smooth reintegration of microrebooted components, inter-component interactions in a crash-only system ideally use timeouts and, if no response is received to a call within the allotted time frame, the caller can gracefully recover. Such timeouts provide an orthogonal mechanism for turning non-Byzantine failures into fail-stop events, which are easier to accommodate and contain. When a component invokes a currently microrebooting component, it receives a `RetryAfter(t)` exception; the call can then be re-issued after the estimated recovery time $t$, if it is idempotent. For non-idempotent calls, rollback or compensating operations can be used. If components transparently recover in-flight requests this way, intra-system component failures and microreboots can be hidden from end users.

**Leases:** Resources in a frequently-microrebooting system should be leased, to improve the reliability of cleaning up after μRBs, which may otherwise leak resources. In addition to memory and file descriptors, we believe certain types of persistent state should carry long-term leases; after expiration, this state can be deleted or archived out of the system. CPU execution time should also be leased: if a computation hangs and does not renew its execution lease, it should be terminated with a μRB. If requests can carry a time-to-live, then stuck requests can be automatically purged from the system once this TTL runs out.

The crash-only design approach embodies well-known principles for robust programming of distributed systems. We push these principles to finer levels of granularity within applications, giving non-distributed applications the robustness of their distributed brethren. In the next section we describe the application of some of these design principles to the implementation of a platform for microrebootable applications.

# 3 A Microrebootable Prototype

The enterprise edition of Java (J2EE) [40] is a framework for building large-scale Internet services. Motivated by its frequent use for critical Internet-connected applications (e.g., 40% of the current enterprise application market [3]), we chose to add microreboot capabilities to an open-source J2EE application server (JBoss [21]) and converted a J2EE application (RUBiS [37]) to the crash-only model. The changes we made to the JBoss platform universally benefit all J2EE applications running on it. In this section we describe the details of J2EE and our prototype.

## 3.1 The J2EE Component Framework

A common design pattern for Internet applications is the three-tiered architecture: the presentation tier consists of stateless Web servers, the application tier runs the application per se, and the persistence tier stores long-term data in one or more databases. J2EE is a framework designed to simplify developing applications for this model.

J2EE applications consist of portable components, called Enterprise Java Beans (EJBs), and platform-specific XML deployment descriptor files. A J2EE application server, akin to an operating system for Internet services, uses the deployment information to instantiate an application's EJBs inside management containers; there is one container per EJB object, and it manages all instances of that object. The server-managed containers provide the application components with a rich set of services: thread pooling and lifecycle management, client session management, database connection pooling, transaction management, security and access control, etc. In theory, a J2EE application should be able to run on any J2EE application server, with modifications only needed in the deployment descriptors.

End users interact with a J2EE application through a Web interface, the application's presentation tier, encapsulated in a WAR – Web ARchive. The WAR component consists of servlets and Java Server Pages (JSPs) hosted in a Web server; they invoke methods on the EJBs and then format the returned results for presentation to the end user. Invoked EJBs can call on other EJBs, interact with the back-end databases, invoke other Web services, etc.

An EJB is similar to an event handler, in that it does not constitute a separate locus of control – a single Java thread shepherds a user request through multiple EJBs, from the point it enters the application tier until it returns to the Web tier. EJBs provide a level of componentization that is suitable for building crash-only applications.

## 3.2 Microreboot Machinery

We added a microreboot method to JBoss that can be invoked programatically from within the server, or remotely, over HTTP. Since we modified the JBoss server, microreboots can now be performed on any J2EE application; however, this is safe only if the application is crash-only. The microreboot method can be applied to one or more EJB or WAR components. It destroys all extant instances of the corresponding objects, kills all shepherding threads associated with those instances, releases all associated resources, discards server metadata maintained on behalf of the component(s), and then reinstantiates and initializes the component(s).

The only resource we do not discard on a μRB is the component's classloader. JBoss uses a separate class loader for each EJB to provide appropriate sandboxing between components; when a caller invokes an EJB method, the caller's thread switches to the EJB's classloader. A Java class' identity is determined both by its name and the classloader responsible for loading it; discarding an EJB's classloader upon μRB would unnecessarily complicate the update of internal references to the microrebooted component. Preserving the classloader does not violate any of the sandboxing properties. Keeping the classloader active does not reinitialize EJB static variables upon μRB, but this is acceptable, since J2EE strongly discourages the use of mutable static variables anyway, as this would prevent transparent replication of EJBs in clusters.

Some EJBs cannot be microrebooted individually, because EJBs might maintain references to other EJBs and because certain metadata relationships can span containers. Thus, whenever an EJB is microrebooted, we microreboot the transitive closure of its inter-EJB dependents as a group. To determine these recovery groups, we examine the EJB deployment descriptors; the information on references is typically used by J2EE application servers to determine the order in which EJBs should be deployed.

## 3.3 A Crash-Only Application

Although many companies use JBoss to run their production applications, we found them unwilling to share their applications with us. Instead, we converted Rice University's RUBiS [37], a J2EE/Web-based auction system that mimics eBay's functionality, into eBid – a crash-only version of RUBiS with some additional functionality. eBid maintains user accounts, allows bidding on, selling, and buying of items, has item search facilities, customized information summary screens, user feedback pages, etc.

**State segregation**: E-commerce applications typically handle three types of important state: long-term data that must persist for years (such as customer account activity), session data that needs to persist for the duration of a user session (e.g., shopping carts or workflow state in enterprise applications), and static presentation data (GIFs, HTML, JSPs, etc.). eBid keeps these types of state in a database, dedicated session state storage, and an Ext3FS filesystem (optionally mounted read-only), respectively.

eBid uses only two types of EJBs: entity EJBs and stateless session EJBs. An entity EJB implements a persistent application object, in the traditional OOP sense, with each instance's state mapped to a row in a database table. Stateless session EJBs are used to perform higher level operations on entity EJBs: each end user operation is implemented by a stateless session EJB interacting with several entity EJBs. For example, there is a "place bid on item X" EJB that interacts with entity EJBs User, Item, and Bid. This mixed OO/procedural design is consistent with best practices for building scalable J2EE applications [10].

*Persistent state* in eBid consists of user account information, item information, bid/buy/sell activity, etc. and is maintained in a MySQL database through 9 entity EJBs: IDManager, User, Item, Bid, Buy, Category, OldItem, Region, and UserFeedback. MySQL is crash-safe and recovers fast for our datasets (132K items, 1.5M bids, 10K users). Each entity bean uses container-managed persistence, a J2EE mechanism that delegates management of entity data to the EJB's container. This way, JBoss can provide relatively transparent data persistence, relieving the programmer from the burden of managing this data directly or writing SQL code to interact with the database. If an EJB is involved in any transactions at the time of a microreboot, they are all automatically aborted by the container and rolled back by the database.

*Session state* in eBid takes the form of items that a user selects for buying/selling/biddding, her userID, etc. Such state must persist on the application server for long enough to synthesize a user session from independent stateless HTTP requests, but can be discarded when the user logs out or the session times out. Users are identified using HTTP cookies. Many commercial J2EE application servers store session state in middle tier memory, in which case a server crash or EJB microreboot would cause the corresponding user sessions to be lost. In our prototype, to ensure the session state survives μRBs, we keep it outside the application in a dedicated session state repository.

We have two options for session state storage. First, we built FastS, an in-memory repository inside JBoss's embedded Web server. The API consists of methods for reading/writing `HttpSession` objects atomically. FastS illustrates how session state can be segregated from the application, yet still be kept within the same Java virtual machine (JVM). Isolated behind compiler-enforced barriers, FastS provides fast access to session objects, but only survives μRBs. Second, we modified SSM [26], a clustered session state store with a similar API to FastS. SSM maintains its state on separate machines; isolated by physical barriers, it provides slower access to session state, but survives μRBs, JVM restarts, as well as node reboots. The session storage model is based on leases, so orphaned session state is garbage-collected automatically.

**Isolation and decoupling**: Compiler-enforced interfaces and type safety provide operational isolation between EJBs. EJBs cannot name each others' internal variables, nor are they allowed to use mutable static variables. EJBs obtain references to each other (in order to make inter-EJB method calls) from a naming service (JNDI) provided by JBoss; references may be cached once obtained. The inter-EJB calls themselves are mediated by the application server via the containers and a suite of interceptors, in order to abstract away details of remote invocation and replication in the cases when EJBs are replicated for performance or load balancing reasons.

Besides preservation of state across microreboots, the segregation of session state in eBid offers recovery decoupling as well, since data shared across components by means of a state store frees the components from having to be recovered together. Such segregation also helps to quickly reintegrate recovered components, because they do not need to perform data recovery following a μRB.

## 4 Evaluation Framework

To evaluate our prototype, we developed a client emulator, a fault injector, and a system for automated failure detection, diagnosis, and recovery. We injected faults in eBid and measured the recovery properties of microrebooting.

We wrote a **client emulator** using some of the logic in the load generator shipped with RUBiS. Human clients are modeled using a Markov chain with 25 states corresponding to the various end user operations possible in eBid, such as *Login*, *BuyNow*, or *AboutMe*; transitioning to a state causes the client to issue a corresponding HTTP request. Inbetween successive "URL clicks," emulated clients have independent think times based on an exponential random distribution with a mean of 7 seconds and a maximum of 70 seconds, as in the TPC-W benchmark [38]. We chose transition probabilities representative of online auction users; the resulting workload, shown in Table 1, mimics the real workload seen by a major Internet auction site [16].

| User operation results mostly in... | % of all requests |
|---|---|
| Read-only DB access (e.g., browse a category) | 32% |
| Initialization/deletion of session state (e.g., login) | 23% |
| Exclusively static HTML content (e.g., home page) | 12% |
| Search (e.g., search for items by name) | 12% |
| Session state updates (e.g., select item for bid) | 11% |
| Database updates (e.g., leave seller feedback) | 10% |

Table 1: Client workload used in evaluating microreboot-based recovery.

To enable automatic recovery, we implemented **failure detection** in the client emulator and placed primitive diagnosis facilities in an external recovery manager. While real end users' Web browsers certainly do not report failures to the Internet services they use, our client-side detection mimics WAN services that deploy "client-like" end-to-end monitors around the Internet to detect a service's user-visible failures [22]. Such a setup allows our measurements to focus on the recovery aspects of our prototype, rather than the orthogonal problem of detection and diagnosis.

We implemented two fault detectors. The first one is simple and fast: if a client encounters a network-level error (e.g., cannot connect to server) or an HTTP 4xx or 5xx error, then it flags the response as faulty. If no such errors occur, the received HTML is searched for keywords indicative of failure (e.g., "exception," "failed," "error"). Finally, the detection of an application-specific problem can also mark the response as faulty (such problems include being prompted to log in when already logged in, encountering negative item IDs in the reply HTML, etc.)

The second fault detector submits in parallel each request to the application instance we are injecting faults into, as well as to a separate, known-good instance on another machine. It then compares the result of the former to the "truth" provided by the latter, flagging any differences as failures. This detector is the only one able to identify complex failures, such as the surreptitious corruption of the dollar amount in a bid. Certain tweaks were required to account for timing-related nondeterminism.

We built a recovery manager (*RM*) that performs simple **failure diagnosis** and **recovers** by: microrebooting EJBs, the WAR, or all of eBid; restarting the JVM that runs JBoss (and thus eBid as well); or rebooting the operating system. *RM* listens on a UDP port for failure reports from the monitors, containing the failed URL and the type of failure observed. Using static analysis, we derived a mapping from each eBid URL prefix to a path/sequence of calls between servlets and EJBs. The recovery manager maintains for each component in the system a score, which gets incremented every time the component is in the path originating at a failed URL. *RM* decides what and when to (micro)reboot based on hand-tuned thresholds. Accurate or sophisticated failure detection was not the topic of this work; our simplistic approach to diagnosis often yields false positives, but part of our goal is to show that even the mistakes resulting from simple or "sloppy" diagnosis are tolerable because of the very low cost of μRBs.

The recovery manager uses a simple **recursive recovery policy** [8] based on the principle of trying the cheapest recovery first. If this does not help, *RM* reboots progressively larger subsets of components. Thus, *RM* first microreboots EJBs, then eBid's WAR, then the entire eBid application, then the JVM running the JBoss application server, and finally reboots the OS; if none of these actions cure the failure symptoms, *RM* notifies a human administrator. In order to avoid endless cycles of rebooting, *RM* also notifies a human whenever it notices recurring failure patterns. The recovery action per se is performed by remotely invoking JBoss's microreboot method (for EJB, WAR, and eBid) or by executing commands, such as `kill -9`, over `ssh` (for JBoss and node-level reboot).

We evaluated the availability of our prototype using a new metric, **action-weighted throughput** ($T_{\text{aw}}$). We view a user *session* as beginning with a login operation and ending with an explicit logout or abandonment of the site. Each session consists of a sequence of user *actions*. Each user action is a sequence of *operations* (HTTP requests) that culminates with a "commit point": an operation that must succeed for that user action to be considered successful as a whole (e.g., the last operation in the action of placing a bid results in committing that bid to the database).

An action succeeds or fails atomically: if all operations within the action succeed, they count toward action-weighted goodput ("good $T_{\text{aw}}$"); if an operation fails, all operations in the corresponding action are marked failed, counting toward action-weighted badput ("bad $T_{\text{aw}}$"). Unlike simple throughput, $T_{\text{aw}}$ accounts for the fact that both long-running and short-running operations must succeed for a user to be happy with the service. $T_{\text{aw}}$ also captures the fact that, when an action with many operations succeeds, it generally means the user did more work than in a short action. Figure 1 gives an example of how we use $T_{\text{aw}}$ to compare recovery by μRB to recovery by JVM restart.

## 5 Evaluation Results

We used our prototype to answer four questions about microrebooting: Are μRBs effective in recovering from failures? Are μRBs any better than JVM restarts? Are μRBs useful in clusters? Do μRB-friendly architectures incur a performance overhead? Section 6 will build upon these results to show how microrebooting can change the way we manage failures in Internet services.

We used 3GHz Pentium machines with 1GB RAM for Web and middle tier nodes; databases were hosted on Athlon 2600xp+ machines with 1.5 GB of RAM and 7200rpm 120GB disks; emulated clients ran on a variety of multiprocessor machines. All machines were interconnected by a 100 Mbps Ethernet switch and ran Linux kernel 2.6.5 with Sun Java 1.4.1 and Sun J2EE 1.3.1.

### 5.1 Is Microrebooting Effective?

Despite J2EE's popularity, we were unable to find any published systematic studies of faults occurring in production J2EE systems. In deciding what faults to inject in our prototype, we relied on advice from colleagues in industry, who routinely work with enterprise applications or application servers [13, 14, 24, 32, 35, 36]. They found that production J2EE systems are most frequently plagued by deadlocked threads, leak-induced resource exhaustion, bug-induced corruption of volatile metadata, and various Java exceptions that are handled incorrectly.

We therefore added hooks in JBoss for injecting artificial deadlocks, infinite loops, memory leaks, JVM memory exhaustion outside the application, transient Java exceptions to stress eBid's exception handling code, and corruption of various data structures. In addition to these hooks, we also used FIG [6] and FAUmachine [7] to inject low-level faults underneath the JVM layer.

eBid, being a crash-only application, has relatively little volatile state that is subject to loss or corruption – much of the application state is kept in FastS / SSM. We can, however, inject faults in the data handling code, such as the code that generates application-specific primary keys for identifying rows in the DB corresponding to entity bean instances. We also corrupt class attributes of the stateless session beans. In addition to application data, we corrupt metadata maintained by the application server, but accessible to eBid code: the JNDI repository, that maps EJB names to their containers, and the transaction method map stored in each entity EJB's container. Finally, we corrupt data inside the session state stores (via bit flips) and in the database (by manually altering table contents).

We perform three types of data corruption: (a) set a value to *null*, which will generally elicit a `NullPointerException` upon access; (b) set an *invalid* value, i.e., a non-null value that type-checks but is invalid from the application's point of view, such as a userID larger than the maximum userID; and (c) set to a *wrong* value, which is valid from the application's point of view, but incorrect, such as swapping IDs between two users.

After injecting a fault, we used the recursive policy described earlier to recover the system. We relied on our comparison-based failure detector to determine whether a recovery action had been successful or not; when failures were still encountered, recovery was escalated to the next level in the policy. In Table 2 we show the worst-case scenario encountered for each type of injected fault. In reporting the results, we differentiate between *resuscitation*, or restoring the system to a point from which it can resume the serving of requests for all users, without necessarily having fixed the resulting database corruption, and *recovery* – bringing the system to a state where it functions with a 100% correct database. Financial institutions often aim for resuscitation, applying compensating transactions at the end of the business day to repair database inconsistencies [36]. A $\approx$ sign in the rightmost column indicates that additional manual database repair actions were required to achieve correct recovery after resuscitation.

Based on these results, we conclude that EJB-level or WAR-level microrebooting in our J2EE prototype is effective in recovering from the majority of failure modes seen in today's production J2EE systems (first 19 rows of Table 2). Microrebooting is ineffective against other types of failures (last 7 rows), where coarser grained reboots or manual repair are required. Fortunately, these failures do not constitute a significant fraction of failures in real J2EE systems. While certain faults (e.g., JNDI corruption) could certainly be cured with non-reboot approaches, we consider the reboot-based approach simpler, quicker, and more reliable. In the cases where manual actions were required to restore service correctness, a JVM restart presented no benefits over a component μRB.

Rebooting is a common way to recover middleware in the real world, so for the rest of this paper we compare EJB-level microrebooting to JVM process restart, which restarts JBoss and, implicitly, eBid.

## 5.2 Is a Microreboot Better Than a Full Reboot?

With respect to availability, Internet service operators care mostly about how many user requests their system turns away during downtime. We therefore evaluate microrebooting with respect to this end-user-aware metric, as captured by $T_{\text{aw}}$. We inject faults in our prototype and then allow the recovery manager (*RM*) to recover the system automatically in two ways: by restarting the JVM process running JBoss, or by microrebooting one or more EJBs, respectively. Recovery is deemed successful when end users do not experience any more failures after recovery. Figure 1 shows the results of one such experiment, in which we injected three different faults every 10 minutes. Session state is stored in FastS. We ran a load of 500 concurrent clients connected to one application server node; for our specific setup, this lead to a CPU load average of 0.7, which is similar to that seen in deployed Internet systems [29, 15]. Unless otherwise noted, we use 500 concurrent clients per node in each subsequent experiment.

| Injected Fault | Type | Reboot level | + |
|---|---|---|---|
| Deadlock | | EJB | |
| Infinite loop | | EJB | |
| Application memory leak | | EJB | |
| Transient exception | | EJB | |
| Corrupt primary keys | set null | EJB | |
| | invalid | EJB | |
| | wrong | EJB | $\approx$ |
| Corrupt JNDI entries | set null | EJB | |
| | invalid | EJB | |
| | wrong | EJB | |
| Corrupt transaction method map | set null | EJB | |
| | invalid | EJB | |
| | wrong | EJB | $\approx$ |
| Corrupt stateless session EJB attributes | set null | unnecessary | |
| | invalid | unnecessary | |
| | wrong | EJB+WAR | $\approx$ |
| Corrupt data inside FastS | set null | WAR | |
| | invalid | WAR | |
| | wrong | WAR | $\approx$ |
| Corrupt data inside SSM | corruption detected via checksum; bad object automatically discarded | | |
| Corrupt data inside MySQL | database table repair needed | | |
| Memory leak outside application | intra-JVM | JVM/JBoss | |
| | extra-JVM | OS kernel | |
| Bit flips in process memory | | JVM/JBoss | $\approx$ |
| Bit flips in process registers | | JVM/JBoss | $\approx$ |
| Bad system call return values | | JVM/JBoss | |

Table 2: Recovery from injected faults: worst case scenarios. Aside from EJB, JBoss, and operating system reboots, some faults require microrebooting eBid's Web component (WAR). In two cases no resuscitation is needed, because the injected fault is "naturally" expunged from the system after the first call fails. In the case of recovering persistent data, this is either done automatically (transaction rollback), or, in the case of injecting *wrong* data, manual reconstruction of the data in the DB is often required (indicated by $\approx$ in the last column). We used the comparison-based fault detector for all experiments in this table.

Overall, using μRBs instead of JVM restarts reduced the number of failed requests by 98%. Visually, the impact of a failure and recovery event can be estimated by the area of the corresponding dip in good $T_{\text{aw}}$, with larger dips indicating higher service disruption. The area of a $T_{\text{aw}}$ dip is determined by its width (i.e., time to recover) and depth (i.e., the throughput of requests turned away during recovery). We now consider each factor in isolation.

**Microreboots recover faster.** The wider the dip in $T_{\text{aw}}$, the more requests arrive during recovery; since these requests fail, they cause the corresponding user actions to fail, thus retroactively marking the actions' requests as failed. We measured recovery time at various granularities and summarize the results in Table 3. In the two right columns we break down recovery time into how long the target takes to crash (be forcefully shut down) and how long it takes to reinitialize. EJBs recover an order of magnitude faster than JVM restart, which explains why the width of the good $T_{aw}$ dip in the μRB case is negligible.

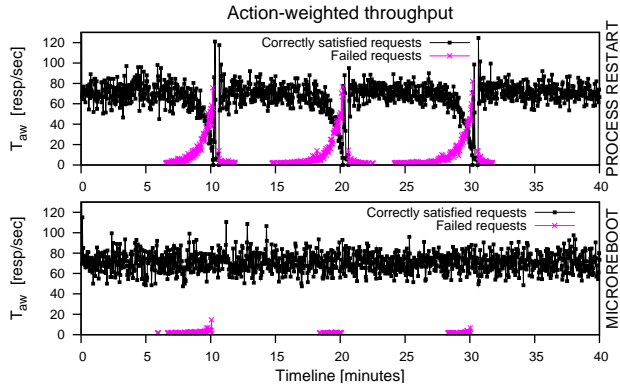

Figure 1: Using $T_{aw}$ to compare JVM process restart to EJB microreboot. Each sample point represents the number of successful (failed) requests observed during the corresponding second. At $t=10$ min, we corrupt the transaction method map for *EntityGroup*, the EJB recovery group that takes the longest to recover. At $t=20$ min, we corrupt the JNDI entry for RegisterNewUser, the next-slowest in recovery. At $t=30$ min, we inject a transient exception in BrowseCategories, the entry point for all browsing (thus, the most-frequently called EJB in our workload). Overall, 11,752 requests (3,101 actions) failed when recovering with a process restart, shown in the top graph; 233 requests (34 actions) failed when recovering by microrebooting one or more EJBs. Thus, the average is 3,917 failed requests (1,034 actions) per process restart, and 78 failed requests (11 actions) per microreboot of one or more EJBs.

| Component name | μRB time (msec) | Crash (msec) | Reinit (msec) |
|---|---|---|---|
| AboutMe | 551 | 9 | 542 |
| Authenticate | 491 | 12 | 479 |
| BrowseCategories | 411 | 11 | 400 |
| BrowseRegions | 416 | 15 | 401 |
| BuyNow* | 471 | 9 | 462 |
| CommitBid | 533 | 8 | 525 |
| CommitBuyNow | 471 | 9 | 462 |
| CommitUserFeedback | 531 | 9 | 522 |
| DoBuyNow | 427 | 10 | 417 |
| EntityGroup* | 825 | 36 | 789 |
| IdentityManager* | 461 | 10 | 451 |
| LeaveUserFeedback | 484 | 10 | 474 |
| MakeBid | 514 | 9 | 515 |
| OldItem* | 529 | 10 | 519 |
| RegisterNewItem | 447 | 13 | 434 |
| RegisterNewUser | 601 | 13 | 588 |
| SearchItemsByCategory | 442 | 14 | 428 |
| SearchItemsByRegion | 572 | 8 | 564 |
| UserFeedback* | 483 | 11 | 472 |
| ViewBidHistory | 507 | 11 | 496 |
| ViewUserInfo | 415 | 10 | 405 |
| ViewItem | 446 | 10 | 436 |
| WAR (Web component) | 1,028 | 71 | 957 |
| Entire eBid application | 7,699 | 33 | 7,666 |
| JVM/JBoss process restart | 19,083 | $\approx 0$ | $\approx 19,083$ |

Table 3: Average recovery times under load, in msec, for the individual components, the entire application, and the JVM/JBoss process. EJBs with a * superscript are entity EJBs, while the rest are stateless session EJBs. Averages are computed across 10 trials per component, on a single-node system under sustained load from 500 concurrent clients. Recovery for individual EJBs ranges from 411-601 msec.

As described in Section 3, some EJBs have interdependencies, captured in deployment descriptors, that require them to be microrebooted together. eBid has one such recovery group, EntityGroup, containing 5 entity EJBs: Category, Region, User, Item, and Bid – any time one of these EJBs requires a μRB, we microreboot the entire EntityGroup. Restarting the entire eBid application is optimized to avoid restarting each individual EJB, which is why eBid takes less than the sum of all components to crash and start up. For the JVM crash, we use operating system-level `kill -9`.

All reboot-based recovery times are dominated by initialization. In the case of JVM-level restart, 56% of the time is spent initializing JBoss and its more than 70 services (transaction service takes 2 sec to initialize, embedded Web server 1.8 sec, JBoss's control & management service takes 1.2 sec, etc.). Most of the remaining 44% startup time is spent deploying and initializing eBid's EJBs and WAR. For each EJB, the deployer service verifies that the EJB object conforms to the EJB specification (e.g., has the required interfaces), then it allocates and initializes a container, sets up an object instance pool, sets up the security context, inserts an appropriate name-to-EJB mapping in JNDI, etc. Once initialization completes, the individual EJBs' `start()` methods are invoked. Removing an EJB from the system follows a reverse path.

**Microreboots reduce functional disruption** during recovery. Figure 1 shows that good $T_{aw}$ drops all the way to zero during a JVM restart, i.e., the system serves no requests during that time. In the case of microrebooting, though, the system continues serving requests while the faulty component is being recovered. We illustrate this effect in Figure 2, graphing the availability of eBid's functionality as perceived by the emulated clients. We group all eBid end user operations into 4 functional groups – Bid/Buy/Sell, Browse/View, Search, and User Account operations – and zoom in on one of the recovery events of Figure 1.

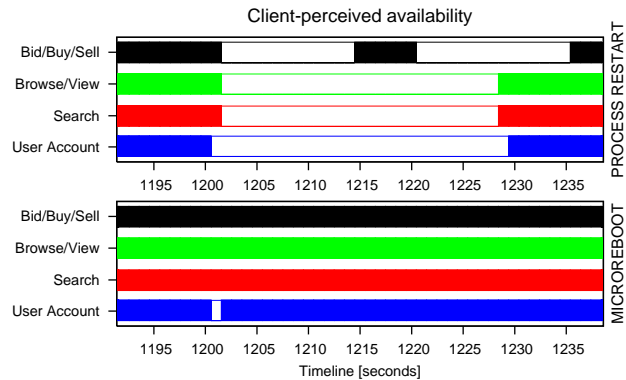

Figure 2: Functional disruption as perceived by end users. For each point $t$ along the horizontal axis, a solid vertical line/bar indicates that, at time $t$, the service was *not* perceived as unavailable by *any* end user. A gap in an interval $[t_1,t_2]$ indicates that some request, whose processing spanned $[t_1,t_2]$ in time, eventually failed, suggesting the site was down.

While the faulty component is being recovered by microrebooting, all operations in other functional groups succeed. Even within the "User Account" group itself, many operations are served successfully during recovery (however, since RegisterNewUser requests fail, we show the entire group as unavailable). Fractional service degradation compounds the benefits of swift recovery, further increasing end user-perceived availability of the service.

**Microreboots reduce lost work.** In Figure 1, a number of requests fail *after* JVM-level recovery has completed; this does not happen in the microreboot case. These failures are due to the session state having been lost during recovery (FastS does not survive JVM restarts). Had we used SSM instead of FastS, the JVM restart case would not have exhibited failed requests following recovery, and a fraction of the retroactively failed requests would have been successful, but the overall good $T_{aw}$ would have been slightly lower (see Section 5.4). Using μRBs in the FastS case allowed the system to both preserve session state across recovery and avoid cross-JVM access penalties.

### 5.3 Is Microrebooting Useful in Clusters?

In a typical Internet cluster, the unit of recovery is a full node, which is small relative to the cluster as a whole. To learn whether μRBs can yield any benefit in such systems, we built a cluster of 8 independent application server nodes. Clusters of 2-4 J2EE servers are typical in enterprise settings, with high-end financial and telecom applications running on 10-24 nodes [15]; a few gigantic services, like eBay's online auction service, run on pools of clusters totaling 2000 application servers [11].

We distribute incoming load among nodes using a client-side load balancer *LB*. Under failure-free operation, *LB* distributes new incoming login requests evenly between the nodes and, for established sessions, *LB* implements session affinity (i.e., non-login requests are directed to the node on which the session was originally established). We inject a μRB-recoverable fault from Table 2 in one of the server instances, say $N_{bad}$; the failure detectors notice failures and report them to the recovery manager. When *RM* decides to perform a recovery, it first notifies *LB*, which redirects requests bound for $N_{bad}$ uniformly to the good nodes; once $N_{bad}$ has recovered, *RM* notifies *LB*, and requests are again distributed as before the failure.

**Failover under normal load**. We first explored the configuration that is most likely to be found in today's systems: session state stored locally at each node; we use FastS. During failover, those requests that do not require session state, such as searching or browsing, will be successfully served by the good nodes; requests that require session state will fail. We injected a fault in the most-frequently called component (BrowseCategories) and ran the experiment in four clusters of different sizes; the load was 500 clients/node.

The left graph in Figure 3 shows the results. When recovering $N_{bad}$ with a JVM restart, the number of user requests that fail is dominated by the number of sessions that were established at the time of recovery on $N_{bad}$. In the case of EJB-level microrebooting, the number of failed requests is roughly proportional to the number of requests that were in flight at the time of recovery or were submitted during recovery. Thus, as the cluster grows, the number of failed user requests stays fairly constant. When recovering with JVM restart, on average 2,280 requests failed; in the case of microrebooting, 162 requests failed.

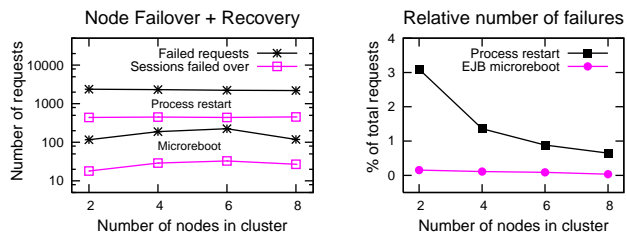

Figure 3: Failover under normal load. On the left we show the number of requests and failed-over sessions for the case of JVM restart and μRB, respectively. On the right we show what fraction of total user requests failed in our test's 10-minute interval, as a function of cluster size.

Although the relative benefit of microrebooting decreases as the number of cluster nodes increases (right graph in Figure 3), recovering with a microreboot will always result in fewer failed requests than a JVM restart, regardless of cluster size or of how many clients each cluster node serves. Thus, it always improves availability. If a cluster aimed for the level of availability offered by today's telephone switches, then it would have to offer six nines of availability, which roughly means it must satisfy 99.9999% of requests it receives (i.e., fail at most 0.0001% of them). Our 8-node cluster served $33.8 \times 10^4$ requests over the course of 10 minutes; extrapolated to a 24-node cluster of application servers, this implies $53.3 \times 10^9$ requests served in a year, of which a six-nines cluster can fail at most $53.3 \times 10^3$. If using JVM restarts, this number allows for 23 single-node failovers during the whole year; if using microreboots, 329 failures would be permissible.

We repeated some of the above experiments using SSM. The availability of session state during recovery was no longer a problem, but the per-node load increased during recovery, because the good nodes had to (temporarily) handle the $N_{bad}$-bound requests. In addition to the increased load, the session state caches had to be populated from SSM with the session state of $N_{bad}$-bound sessions. These factors resulted in an increased response time that often exceeded 8 sec when using JVM restarts; microrebooting was sufficiently fast to make this effect unobservable. Overload situations are mitigated by overprovisioning the cluster, so we investigate below whether microrebooting can reduce the need for additional hardware.

**Microreboots preserve cluster load dynamics**. We repeated the experiments described above using FastS, but doubled the concurrent user population to 1000 clients/node. The load spike we model is very modest compared to what can occur in production systems (e.g., on 9-11, CNN.com faced a 20-fold surge in load, which caused their cluster to collapse under congestion [23]). We also allow the system to stabilize at the higher load prior to injecting faults (for this reason, the experiment's time interval was increased to 13 minutes). Since JVM restarts are more disruptive than microreboots, a mild two-fold change in load and stability in initial conditions favor full process restarts more than µRBs, so the results shown here are conservative with respect to microrebooting. Figure 4 shows that response time was preserved while recovering with µRBs, unlike when using JVM restarts.

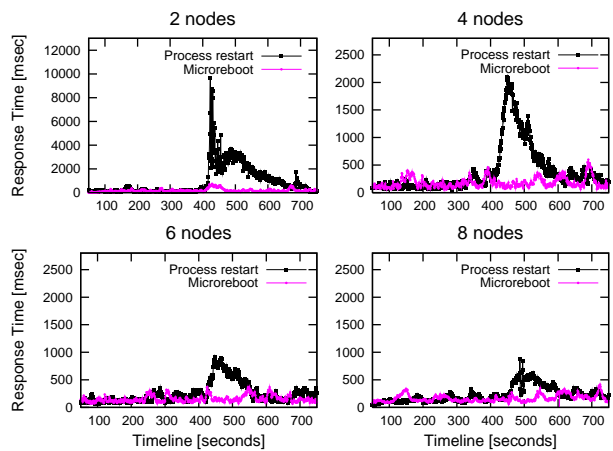

Figure 4: Failover under doubled load. We show average response time per request, computed over 1-second intervals, in 4 different cluster configurations (2,4,6,8 nodes). eBid uses FastS for storing session state, in both the JVM restart and microreboot case. Vertical scales of the four graphs differ, to enhance visibility of details.

Stability of response time results in improved service to the end users. It is known that response times exceeding 8 seconds cause computer users to get distracted from the task they are pursuing and engage in others [31, 4], making this a common threshold for Web site abandonment [44]; not surprisingly, service level agreements at financial institutions often stipulate 8 seconds as a maximum acceptable response time [28]. We therefore measured how many requests exceeded this threshold during failover; Table 4 shows the corresponding results.

| # of nodes | 2 | 4 | 6 | 8 |
|---|---|---|---|---|
| Process restart | 3,227 | 530 | 55 | 9 |
| EJB microreboot | 3 | 0 | 0 | 0 |

Table 4: Requests exceeding 8 sec during failover under doubled load.

We asked our colleagues in industry whether commercial application servers do admission control when overloaded, and were surprised to learn they currently do not [29, 15]. For this reason, cluster operators need to significantly overprovision their clusters and use complex load balancers, tuned by experts, in order to avert overload and oscillation problems. Microreboots reduce the need for overprovisioning or sophisticated load balancing. Since µRBs are more successful at keeping response times below 8 seconds in our prototype, we would expect user experience to be improved in a clustered system that uses microreboot-based recovery instead of process restarts.

### 5.4 Performance Impact

In this section we measure the performance impact our modifications have on steady-state fault-free throughput and latency. We measure the impact of our microreboot-enabling modifications on the application server, by comparing original JBoss 3.2.1 to the microreboot-enabled variant. We also measure the cost of externalizing session state into a remote state store by comparing eBid with FastS to eBid with SSM. Table 5 summarizes the results.

| Configuration | Throughput [req/sec] | Average Latency [msec] |
|---|---|---|
| JBoss + eBid$_{FastS}$ | 72.09 | 15.02 |
| JBoss$_{\mu RB}$ + eBid$_{FastS}$ | 72.42 | 16.08 |
| JBoss + eBid$_{SSM}$ | 71.63 | 28.43 |
| JBoss$_{\mu RB}$ + eBid$_{SSM}$ | 70.86 | 27.69 |

Table 5: Performance comparison: (a) original JBoss vs. microreboot-enabled JBoss$_{\mu RB}$; (b) intra-JVM session state storage (eBid$_{FastS}$) vs. extra-JVM session state storage (eBid$_{SSM}$).

Throughput varies less than 2% between the various configurations, which is within the margin of error. Latency, however, increases by 70-90% when using SSM, because moving state between JBoss and a remote session state store requires the session object to be marshalled, sent over the network, then unmarshalled; this consumes more CPU than if the object were kept inside the JVM. Since minimum human-perceptible delay is about 100 msec [31], we believe the increase in latency is of little consequence for an interactive Internet service like ours. Latency-critical applications can use FastS instead of SSM. The performance results are within the range of measurements done at a major Internet auction service, where latencies average 33-300 msec, depending on operation, and average throughput is 41 req/sec per node [16].

It is not meaningful to compare the performance of eBid to that of original RUBiS, because the semantics of the applications are different. For example, RUBiS requires users to provide a username and password *each time* they perform an operation requiring authentication. In eBid, users log in once at the beginning of their session; they are subsequently identified based on the HTTP cookies they supply to the server on every access. We refer the reader to [10] for a detailed comparison of performance and scalability for various architectures in J2EE applications.

## 6 A New Approach to Failure Management

The previous section showed microreboots to have significant quantitative benefits in terms of recovery time, functionality disruption, amount of lost work, and preservation of load dynamics in clusters. These quantitative improvements beget a qualitative change in the way we can manage failures in large-scale componentized systems; here we present some of these new possibilities.

### 6.1 Alternative Failover Schemes

In a microrebootable cluster, μRB-based recovery should always be attempted first, prior to failover. As seen earlier, node failover can be destabilizing. In the first set of experiments in Section 5.3, failing requests over to good nodes while $N_{bad}$ was recovering by μRB resulted in 162 failed requests. In Figure 1, however, the average number of failures when requests continued being sent to the recovering node was 78. This shows that μRB without failover improves user-perceived availability over failover and μRB.

The benefit of pre-failover μRB is due to the mismatch between node-level failover and component-level recovery. Coarse-grained failover prevents $N_{bad}$ from serving a large fraction of the requests it could serve while recovering (Figure 2). Redirecting those requests to other nodes will cause many requests to fail (if not using SSM), or at best will unnecessarily overload the good nodes (if using SSM). Should the pre-failover μRB prove ineffective, the load balancer can do failover and have $N_{bad}$ rebooted; the cost of microrebooting in a non-μRB-curable case is negligible compared to the overall impact of recovery.

Using the average of 78 failed requests per microreboot instead of 162, we can update the computation for six-nines availability from Section 5.3. Thus, if using microreboots and *no failover*, a 24-node cluster could fail 683 times per year and still offer six nines of availability. We believe writing microrebootable software that is allowed to fail almost twice every day (683 times/year) is easier than writing software that is not allowed to fail more than once every 2 weeks ($\approx$23 times/year for JVM restart recovery).

Another way to mitigate the coarseness of node-level failover is to use component-level failover; having reduced the cost of a reboot by making it finer-grain, *microfailover* seems a natural solution. Load balancers would have to be augmented with the ability to fail over only those requests that would touch the component(s) known to be recovering. There is no use in failing over any other requests. Microfailover accompanied by microreboot can reduce recovery-induced failures even further. Microfailover, however, requires the load balancer to have a thorough understanding of application dependencies, which might make it impractical for real Internet services.

### 6.2 User-Transparent Recovery

If recovery is sufficiently non-intrusive, then we can use low-level retry mechanisms to hide failure and recovery from callers – if it is brief, they won't notice. Fortunately, the HTTP/1.1 specification [18] offers return code 503 for indicating that a Web server is temporarily unable to handle a request (typically due to overload or maintenance). This code is accompanied by a `Retry-After` header containing the time after which the Web client can retry.

We implemented call retry in our prototype. Previously, the first step in microrebooting a component was the removal of its name binding from JNDI; instead, we bind the component's name to a sentinel during μRB. If, while processing an idempotent request, a servlet encounters the sentinel on an EJB name lookup, the servlet container automatically replies with [`Retry-After 2 seconds`] to the client. We associated idempotency information with URL prefixes based on our understanding of eBid, but this could also be inferred using static call analysis. We measured the effect of HTTP/1.1 retry on calls to four different components, and found that transparent retry masked roughly half of the failures (Table 6); this corresponds to a two-fold increase in perceived availability.

| Operation / component name | No retry | Retry | Delay & retry |
|---|---|---|---|
| ViewItem | 23 | 16 | 8 |
| BrowseCategories | 20 | 8 | 0 |
| SearchItemsByCategory | 31 | 15 | 0 |
| Authenticate | 20 | 9 | 1 |

Table 6: Masking microreboots with HTTP/1.1 `Retry-After`. The data is averaged across 10 trials for each component shown.

The failed requests visible to end users were requests that had already entered the system when the microreboot started. To further reduce failures, we experimented with introducing a 200-msec delay between the sentinel rebind and beginning of the microreboot; this allowed some of the requests that were being processed by the about-to-be-microrebooted component to complete. Of course, a component that has encountered a failure might not be able to process requests prior to recovery, unless only some instances of the EJB are faulty, while other instances are OK (a microreboot recycles all instances of that component). The last column in Table 6 shows a significant further reduction in failed requests. We did not analyze the tradeoff between number of saved requests and the 200-msec increase in recovery time.

### 6.3 Tolerating Lax Failure Detection

In general, downtime for an incident is the sum of the time to detect the failure ($T_{det}$), the time to diagnose the faulty component, and the time to recover. A failure monitor's quality is generally characterized by how quick it is (i.e., $T_{det}$), how many of its detections are mistaken (false positive rate $FP_{det}$), and how many real failures it misses (false negative rate $FN_{det}$). Monitors make tradeoffs between these parameters; e.g., a longer $T_{det}$ generally yields lower $FP_{det}$ and $FN_{det}$, since more sample points can be gathered and their analysis can be more thorough.

Cheap recovery relaxes the task of failure detection in at least two ways. First, it allows for longer $T_{det}$, since the additional requests failing while detection is under way can be compensated for with the reduction in failed requests during recovery. Second, since false positives result in useless recovery leading to unnecessarily failing requests, cheaper recovery reduces the cost of a false positive, enabling systems to accommodate higher $FP_{det}$. Trading away some $FP_{det}$ and $T_{det}$ may result in a lower false negative rate, which could improve availability.

We illustrate $T_{\text{det}}$ relaxation in the left graph of Figure 5. We inject a fault in the most frequently called EJB and delay recovery by $T_{\text{det}}$ seconds, shown along the horizontal axis; we then perform recovery using either a JVM restart or a microreboot. The dotted line indicates that, with µRB-based recovery, a monitor can take up to 53.5 seconds to detect a failure, while still providing higher user-perceived availability than JVM restarts with immediate detection ($T_{\text{det}} = 0$). The two curves in the graph become asymptotically close for large values of $T_{\text{det}}$, because the number of requests that fail during detection (i.e., due to the delay in recovery) eventually dominate those that fail during recovery itself.

Doing real-time diagnosis instead of recovery has an opportunity cost. In this experiment, 102 requests failed during the first second of waiting; in contrast a microreboot averages 78 failed requests and takes 411-825 msec (Table 3), which suggests that microrebooting *during* diagnosis would result in approximately the same number of failures, but offers the possibility of curing the failure before diagnosis completes.

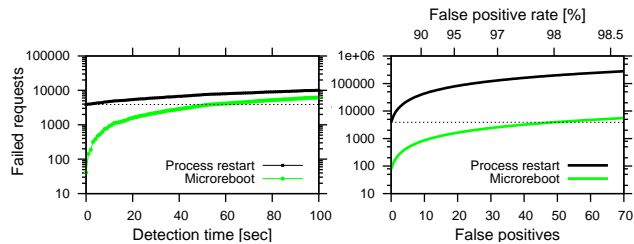

Figure 5: Relaxing failure detection with cheap recovery.

The right graph of Figure 5 shows the effect of false positives on end-user-perceived availability, given the averages from Figure 1: 3,917 failed requests per JVM restart, 78 requests per µRB. False positive detections occur inbetween correct positive detections; the false ones result in pointless recovery-induced downtime, while the correct ones lead to useful recovery. For simplicity, we assume $T_{\text{det}} = 0$. The graph plots the number of failed requests $f(n)$ caused by a sequence of $n$ useless recoveries (triggered by false positives) followed by one useful recovery (in response to the correct positive). A given number $n$ of false positives inbetween successive correct detections corresponds to a $FP_{\text{det}} = n/(n+1)$. The dotted line indicates that the availability achieved with JVM restarts and $FP_{\text{det}} = 0\%$ can be improved with µRB-based recovery even when false positive rates are as high as 98%.

Engineering failure detection that is both fast and accurate is difficult. Microreboots give failure detectors more headroom in terms of detection speed and false positives, allowing them to reduce false negative rates instead, and thus reduce the number of real failures they miss. Lower false negative rates can lead to higher availability. We would expect some of the extra headroom to also be used for improving the precision with which monitors pinpoint faulty components, since microrebooting requires component-level precision, unlike JVM restarts.

### 6.4 Averting Failure with Microrejuvenation

Despite automatic garbage collection, resource leaks are a major problem for many large-scale Java applications; a recent study of IBM customers' J2EE e-business software revealed that production systems frequently crash because of memory leaks [33]. To avoid unpredictable leak-induced crashes, operators resort to preventive rebooting, or software rejuvenation [20]. Some of the largest U.S. financial companies reboot their J2EE servers daily [32] to recover memory, network sockets, file descriptors, etc. In this section we show that µRB-based rejuvenation, or *microrejuvenation*, can be as effective as a JVM restart in preventing leak-induced failures, but cheaper.

We wrote a server-side rejuvenation service that periodically checks the amount of memory available in the JVM; if it drops below $M_{\text{alarm}}$ bytes, then the recovery service microreboots components in a rolling fashion until available memory exceeds a threshold $M_{\text{sufficient}}$; if all EJBs are microrebooted and $M_{\text{sufficient}}$ has not been reached, the whole JVM is restarted. Production systems could monitor a number of additional system parameters, such as number of file descriptors, CPU utilization, lock graphs for identifying deadlocks, etc.

The rejuvenation service does not have any knowledge of *which* components need to be microrebooted in order to reclaim memory. Thus, it builds a list of all components; as components are microrebooted, the service remembers how much memory was released by each one's µRB. The list is kept sorted in descending order by released memory and, the next time memory runs low, the rejuvenation service microrejuvenates components expected to release most memory, re-sorting the list as needed.

We induce memory leaks in two components: ViewItem, a stateless session EJB called frequently in our workload, and Item, an entity EJB part of the long-recovering EntityGroup. We choose leak rates that allow us to keep each experiment under 30 minutes.

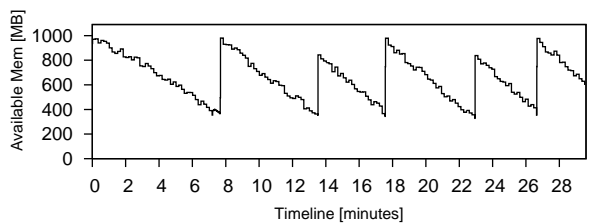

Figure 6: Available memory during microrejuvenation. We inject a 2 KB/invocation leak in Item and a 250 KB/invocation leak in ViewItem. $M_{\text{alarm}}$ is set to 35% of the 1-GByte heap (thus $\approx$ 350 MB) and $M_{\text{sufficient}}$ to 80% ($\approx$ 800 MB).

In Figure 6 we show how free memory varies under a worst-case scenario for microrejuvenation: the initial list of components has the components leaking most memory at the very end. During the first round of microrejuvenation (interval [7.43-7.91] on the timeline), all of eBid ends up rebooted by pieces. During this time, ViewItem is found to have the most leaked memory, and Item the second-most; the list of candidate components is reordered accordingly, improving the efficiency of subsequent rejuvenations. The second time $M_\mathrm{alarm}$ is reached, at $t = 13.8$, microrebooting ViewItem is sufficient to bring available memory above threshold. On the third rejuvenation, both ViewItem and Item require rejuvenation; on the fourth, a ViewItem µRB is again sufficient; and so on.

Repeating the same experiment, but using whole rejuvenation via JVM restarts, resulted in a total of 11,915 requests failed during the 30-minute interval. When microrejuvenating with µRBs, only 1,383 requests failed – an order of magnitude improvement – and good $T_\mathrm{aw}$ never dropped to zero. The commonly used argument to motivate software rejuvenation is that it turns unplanned total downtime into planned total downtime; with microrejuvenation, we can further turn this planned total downtime into planned *partial* downtime.

## 7 Limitations of Recovery by Microreboot

It may appear that µRBs introduce three classes of problems: interruption of a component during a state update, improper reclamation of a microrebooted component's external resources, and delay of a (needed) full reboot.

**Impact on shared state**. If state updates are atomic, as they are with databases, FastS, or SSM, there is no distinction between µRBs and process restarts from the state's perspective. However, the case of non-atomic updates to state shared between components is more challenging: microrebooting one component may leave that state inconsistent, unbeknownst to the other components that share it. A JVM restart, on the other hand, reboots all components simultaneously, so it does not give them an opportunity to see the inconsistent state. J2EE best-practices documents discourage sharing state by passing references between components or using static variables, but we believe this should be a requirement enforced by a suitably modified JIT compiler. Alternatively, if the runtime detects unsafe state sharing practices, it should disable the use of µRBs for the application in question.

Not only does a JVM restart refresh all components, but it also discards the volatile shared state, regardless of whether it is inconsistent or not; µRBs allow that state to persist. In a crash-only system, state that survives the recovery of components resides in a state store that assumes responsibility for data consistency. In order to accomplish this, dedicated state repositories need APIs that are sufficiently high-level to allow the repository to repair the objects it manages, or at the very least to detect corruption. Otherwise, faults and inconsistencies perpetuate; this is why application-generic checkpoint-based recovery in Unix was found not to work well [27]. In the logical limit, all applications become stateless and recovery involves either microrebooting the processing components, or repairing the data in state stores.

**Interaction with external resources**. If a component circumvents JBoss and acquires an external resource that the application server is not aware of, then microrebooting it may leak the resource in a way that a JVM/JBoss restart would not. For example, we experimentally verified that an EJB $X$ can directly open a connection to a database without using JBoss's transaction service, acquire a database lock, then share that connection with another EJB $Y$. If $X$ is microrebooted prior to releasing the lock, $Y$'s reference will keep the database connection open even after $X$'s µRB, and thus $X$'s DB session stays alive. The database will not release the lock until after $X$'s DB session times out. In the case of a JVM restart, however, the resulting termination of the underlying TCP connection by the operating system would cause the immediate termination of the DB session and the release of the lock. If JBoss only knew $X$ acquired a DB session, it could properly free the session even in the case of µRB.

While this example is contrived and violates J2EE programming practices, it illustrates the need for application components to obtain resources exclusively through the facilities provided by their platform.

**Delaying a full reboot**. The more state gets segregated out of the application, the less effective a reboot becomes at scrubbing this data. Moreover, our implementation of µRB does not scrub data maintained by the application server on behalf of the application, such as the database connection pool and various caches. Microreboots also generally cannot recover from problems occurring at layers below the application, such as within the application server or the JVM; these require a full JVM restart instead.

When a full process restart is required, poor failure diagnosis may result in one or more ineffectual component-level µRBs. As discussed in Section 6.3, failure localization needs to be more precise for microreboots than for JVM restarts. Using our recursive policy, microrebooting progressively larger groups of components will eventually restart the JVM, but later than could have been done with better diagnosis. Even in this case, however, µRBs add only a small additional cost to the total recovery cost.

## 8 Generalizing beyond Our Prototype

Some J2EE applications are already microreboot-friendly and require minimal changes to take advantage of our µRB-enabled application server. Based on our experience with other J2EE applications, we learned that the biggest challenges in making them 100% microrebootable are (a) extricating session state handling from the application logic, and (b) ensuring that persistent state is updated with transactions. The rest is already done in our prototype server and can be leveraged across all J2EE applications.

While we feel J2EE makes it easier to write a microrebootable application, because its model is amenable to state externalization and component isolation, we hope to see microreboot support in other types of systems as well. In this section we describe design aspects that deserve consideration in such extensions.

**Isolation:** If there is one property of microrebootable systems that is more critical than all the others, it is the partitioning of the system into fine-grain, well isolated components. While such partitioning is a system-specific task, frameworks like J2EE and .NET [30] can help. Component isolation in J2EE is not enforced by lower-level (hardware) mechanisms, as would be the case with separate process address spaces; consequently, bugs in the Java virtual machine or the application server could result in state corruption crossing component boundaries. Depending on the system, stronger levels of isolation may be warranted, such as can be achieved with processes or virtual machines. Dependencies between components need to be minimized, because a dense dependency graph increases the size of recovery groups, making μRBs take longer and be more disruptive.

**Workload:** Microreboots thrive on workloads consisting of fine-grain, independent requests; if a system is faced with long running operations, then individual components could be periodically microcheckpointed [42] to keep the cost of μRBs low, keeping in mind the associated risk of persistent faults. In the same vein, requests need to be sufficiently self-contained, such that a fresh instance of a microrebooted component can pick up a request and continue processing it where the previous instance left off.

**Resources:** Java does not offer explicit memory release or lease-based allocation, so the best we could do was to call the system garbage collector after μRB. However, this form of resource reclamation does not complete in an amount of time that is independent of the size of the memory, unlike most traditional operating systems. We believe that efficient support for microreboots requires a nearly-constant-time resource reclamation mechanism, to allow microreboots to synchronously clean up resources.

## 9  Related Work

Our work has three major themes: reboot-based recovery, minimizing recovery time, and reducing disruption during recovery. In this section we discuss a small sample of work related to these themes.

Separation of control and data is key to reboot-based recovery. There are many ways to isolate subsystems (e.g., using processes, virtual machines [17], microkernels [25], protection domains [41], etc.). Isolated processing components appeared also in pre-J2EE transaction processing monitors, where each piece of system functionality (e.g., doing I/O with clients, writing to the transaction log) was a separate process communicating with the others using IPC or RPC. Session state was managed in memory by a dedicated component. Although the architecture did not scale very well, the "one component/one process" approach provided better isolation than monolithic architectures and would have been amenable to microrebooting.

Baker [2] observed that emphasizing fast recovery over crash prevention has the potential to improve availability, and she described ways to build distributed file systems such that they recover quickly after crashes. In her design, a "recovery box" safeguards metadata in memory for recovery after a warm reboot. In our work, we provide components for a more general framework that both reduces the impact of a crash and speeds up recovery.

Much work in Internet services has focused on reducing the functional disruption associated with recovering from a transient failure. Failover in clusters is the canonical example; Brewer [5] proposed the "DQ principle" as a way to understand how a partial failure in a multi-node service can be mapped to either a decrease in queries served per second, or a decrease in data returned per query.

Other research systems have embraced the approach of reducing downtime by recovering at sub-system levels. For example, Nooks [41] isolates drivers within lightweight protection domains inside the operating system kernel; when a driver fails, it can be restarted without affecting the rest of the kernel. Farsite [1], a peer-to-peer file system, has been recently restructured as a collection of crash-only components, that are recovered through rebooting. These systems provide examples of microrebootable systems and lend credibility to the belief that non-J2EE systems can be structured for effective microrebootability.

## 10  Conclusions

Employing reboot-based recovery does not mean that the root causes of failures should not be identified and fixed. Rebooting simply provides a separation of concerns between diagnosis and recovery, consistent with the observation that the former is not always a prerequisite for the latter. Moreover, attempting to recover a reboot-curable failure by anything other than a reboot entails the risk of taking longer and being more disruptive than a reboot would have been in the first place, thus hurting availability.

By completely separating process recovery from data recovery, and delegating the latter to specialized state stores, we enabled the use of microreboots to achieve process recovery. In our experiments, microreboots cured the majority of failures that were empirically observed to cause downtime in deployed Internet services. Compared to process restart-based recovery, microrebooting is an order of magnitude faster and less disruptive, even in multi-node clusters.

Regardless of fault, in microrebootable systems one should first attempt microreboot-based recovery: it does not take long and costs very little. Skipping node failover in clusters and microrebooting the faulty node can improve availability over the commonly-used "fail over and reboot node" approach. Microreboot-based recovery can achieve higher levels of availability even when false positive rates in fault detection are as high as 98%. Using microreboots, we were able to reclaim memory leaks in our prototype application without shutting it down, improving availability by an order of magnitude.

There is a significant limitation in developing bug-free software beyond a certain size. Accepting bugs as a fact, we argue that structuring systems for cheap reboot-based recovery provides a promising path toward dependable large-scale software.

## Acknowledgments

We would like to thank David Cheriton and our colleagues in the Recovery-Oriented Computing project for early feedback on this work. We are indebted to our shepherd Jason Nieh, the anonymous OSDI reviewers, and Katerina Argyraki, Kim Keeton, Adam Messinger, Martin Rinard, and Westley Weimer for patiently helping us improve this paper.